\documentclass[11pt, a4paper]{article}
\usepackage{moriond,epsfig}
\def\be{\begin{equation}}
\def\ee{\end{equation}}
\def\bea{\begin{eqnarray}}
\def\eea{\end{eqnarray}}

\begin{document}
\vspace*{4cm}
\title{Beauty and Charm Production Measurements at the Tevatron}
% Changed 040623 - All authors go here
\author{ Thorsten Kuhl}
\address{Universit\"at Mainz, Institut f\"ur Physik, Staudingerweg 7, 55099 Mainz } 

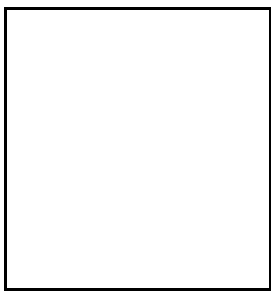
\maketitle\abstracts{
Results for the production of charm and beauty quarks in $\mathrm{p}\overline{\mathrm{p}}$ collisions at $\sqrt{s}=1.96\,\mathrm{TeV}$ (Tevatron) with the two 
multi purpose experiments CDF and D0 using an integrated luminosity of up to $1\,\mathrm{fb^{-1}}$ are presented. With the data measurement of the production mechanism for charm and beauty quarks are done. These measurements are leading to a better understanding of the QCD in the transition region between pertubative and nonpertubative QCD. The charm-charm angular correlation is measured with D mesons at the CDF experiment. The Quarkonium production for charm and beauty final state at D0 and CDF is discussed. 
}

\section{Introduction}
The production of long lived heavy quarks, the charm (c) and beauty (b) quark, in hadron-hadron collisions is and active field of research in Quantum Chromo Dynamics (QCD). In the theoretical treatment the heavy quark mass provides a scale just at the transition between non-pertubative and pertubative QCD. The measurement of cross sections, angular quark-quark correlations and polarisation improves our understanding of the QCD transition region.  This leads to an excellent program for testing one of the fundamental forces. \\
Charm and beauty quarks are produced in huge numbers in proton anti-proton collision at a center-of-mass energy of $\sqrt{s}=1.96\,\mathrm{GeV}$ at the Tevatron collider at Fermilab. CDF~\cite{CDFdetector}  and D0~\cite{D0detector} are the two multi purpose detectors at Fermilab. The two experiments have partly complementary utilities for measurements of heavy quark final states. 
%The CDF experiments has a large tracking volume and a high trigger bandwidth for track based triggers. This makes it to a excellent tool to select fully hadronic final state in charm and beauty physics. 
The large angular coverage for muons and the high muon trigger bandwidth of the D0 detector lead to an excellent efficient selection specially for di-muon hadronic final states like the Upsilon. \\   
The measurements of the charm and beauty cross sections were made with small fraction of the avaible data. For thecharm cross section the CDF experiment showed that using a luminosity of about $6\,\mathrm{pb^{-1}}$ the measured value to be about a factor of two above the theoretical expectation. This is still well inside the scale uncertainty of the theory~\cite{Dxsect}. Also new and updated measurements by both collaborations with data using a luminosity up to $1\,\mathrm{fb^{-1}}$ for the beauty cross sections show the same tendency: the measured cross sections are above the central values of the improved theoretical expectation but well covered inside the theoretical uncertainties~\cite{CDFB,D0B}.\\
%Because of the large theoretical uncertainties for total and differential cross sections, the measurement of the angular correlation for associated charm Mesons and Quarkonium polarisation measurements are done to learn more about the underlaying production mechanism.

\begin{figure}
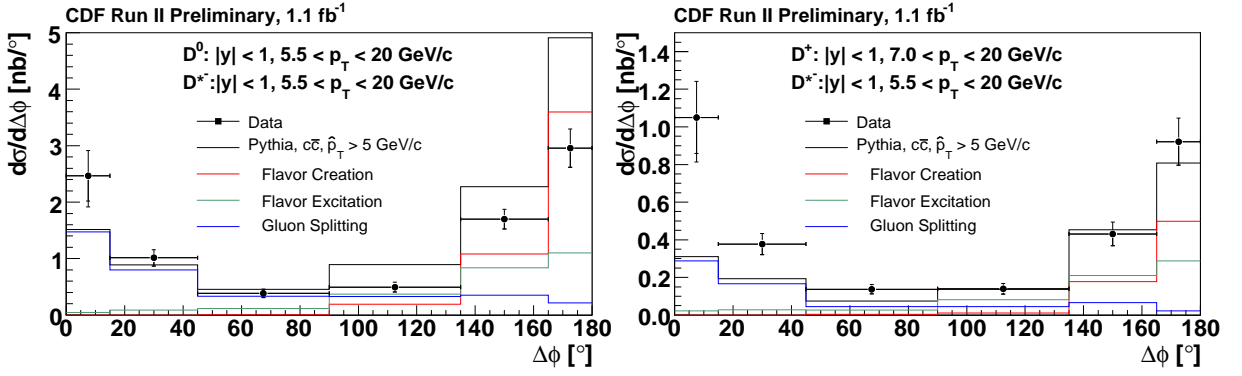

%\rule{5cm}{0.2mm}\hfill\rule{5cm}{0.2mm}
%\vskip 2.5cm
%\rule{5cm}{0.2mm}\hfill\rule{5cm}{0.2mm}
\epsfig{figure=DzeroDstar_xsec_sum.eps, width=0.5\textwidth}\epsfig{figure=DplusDstar_xsec_sum3.eps,width=0.5\textwidth}\\
\caption{  The $D^0D^{*-}$ (left) and $D^+D^{*-}$ (right) pair cross section measured by the CDF experiment as a function of $\Delta \phi$. The measurements are compared to Pythia (black line). Also the different contributions from quark pair production (red), flavour excitation (green) and gluon splitting (blue) are shown.\label{fig:numberone}}
\end{figure}

\section{Charm correlation measurements}
In QCD, the flavour conservation implies, that charm quarks are always produced as quark anti-quark pairs. The azimuthal angle ($\phi$) correlation of the mesons carrying the two heavy quarks in a event gives a possibility to study the underlaying production mechanism in detail~\cite{rfields}. Prompt heavy quark pair production leads to back-to-back production. Heavy quarks produced by splitting of massless gluons have a small $\Delta\phi$. The third production process, the production by flavour excitation leads to a big separation in the rapidity $\eta$ and most of the time one of the two quarks is produced in the forward region. \\
 The CDF collaboration measured the azimute correlation in D-Meson~\cite{CDFcharm} production with an integrated luminosity of $1\,\mathrm{fb^{-1}}$. The angular correlation of more than 2000 events in the two signal modes $D^0D^{*-}$ and $D^+D^{*-}$ was investigated. The $D^0$ and $D^+$ are used for the trigger selection of charm events. The slow pion from the $D^{*-}$ decay leads to a high purity of the samples. Simulation studies have shown that the event selection efficiency factorize into the two single efficiencies of the charm Mesons. Therefore the measured single charm cross sections can be used for the overall normalisation. Figure~\ref{fig:numberone} shows the angular correlation in the $D^0D^{*-}$ and in the $D^+D^{*-}$ channel and compares it to the Pythia simulation. While Pythia gives a fair estimate for the overall production, it leads to a underestimation (overestimation) of thecontribution from pair production (gluon splitting). %The fraction of events produced by flavour excitation is very small because of the limited $\eta$ coverage of the analysis up to $\eta=1$. 
%The biggest systematic uncertainty of about $15\%$ is the estimation of the background from b-decays which was estimated using life time cuts.  

\section{Quarkonium}
\begin{figure}
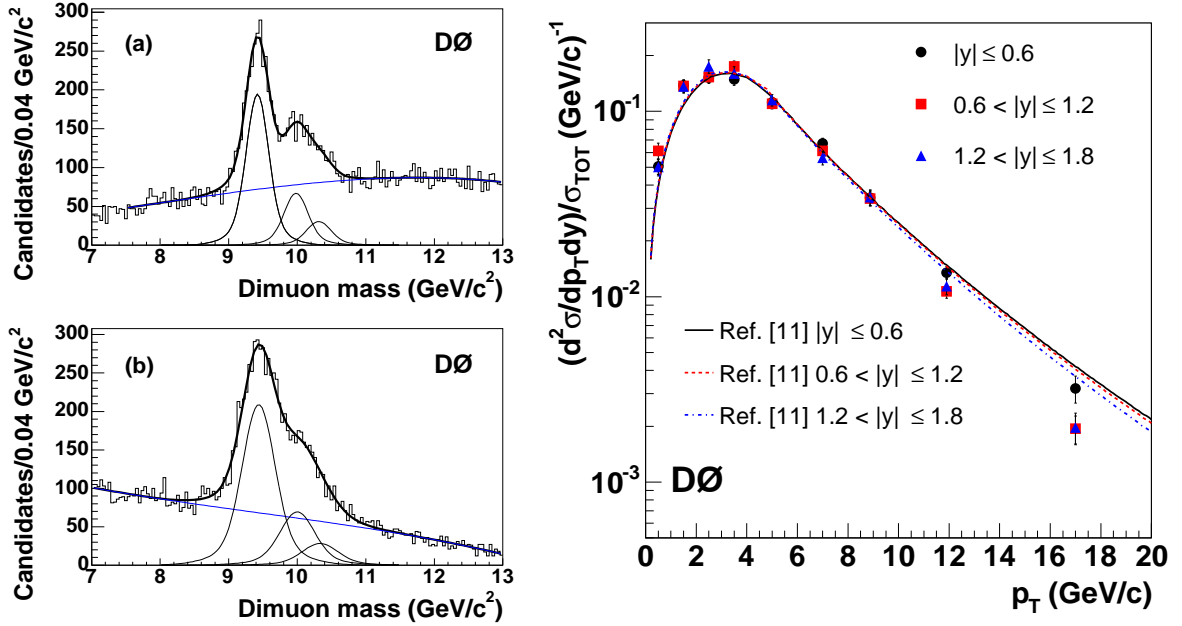

%\rule{5cm}{0.2mm}\hfill\rule{5cm}{0.2mm}
%\vskip 2.5cm
%\rule{5cm}{0.2mm}\hfill\rule{5cm}{0.2mm}
\epsfig{figure=B05AF01.eps, width=0.46\textwidth}\epsfig{figure= B05AF02.eps, width=0.54\textwidth}\\
\caption{  Example of the di-moun spectrum for different bins in the rapidity $y$ (Top: $|y|<0.6$; bottom: $1.2<|y|<1.8$) and $p_T$ between 4 and $6\,\mathrm{GeV}$ (left). The distribution for $\Upsilon(1S)$, $\Upsilon(2S)$ and $\Upsilon(3S)$ are shown as gaussian while a polimon was used for the background. Differential cross section for $\Upsilon(1S)$ production compared with theory prediction for the 3 different $y$ regions  (right).{ \label{fig:numbertwo}}}
\end{figure}

Prior the first measurements at Fermilab, the production of Quarkonium was described by the color singlett model~\cite{colorsing}. However the first measurements at Fermilab showed that this model underestimates the direct production cross section by an order of magnitude for the $J/\Psi$ and by a factor of 50 for the $\Psi(2S)$ production. To address these problems, the color octet model was introduced~\cite{coloroct}. Adjustable hadronisation parameters in this model allow to resonable describe the amplitude and $p_T$-dependence of the production. Recent, different approaches using Pomeronic ideas and $k_T$-factorisation models were introduced to describe the hadronization problem~\cite{khoze,berger}. Both models predict that at sufficient high $p_T$ the $J/\Psi$ and the $\Psi(2S)$ will have a longitudinal polarisation.\\
The D0 experiment selected the $\Upsilon(1S)$,$\Upsilon(2S)$ and $\Upsilon(3S)$ via di-muon decay and measured the $\Upsilon(1S)$ differential cross section~\cite{D0Ups}. Figure~\ref{fig:numbertwo}(left) shows the di-muon mass spectra for two different bins in the pseudorapidity $y$ of the final state. % Three Gaussian are used to estimate the fraction of $\Upsilon(1S)$,  $\Upsilon(2S)$ and  $\Upsilon(3S)$. For the background a polinom was used. 
Clear significant signals of  $\Upsilon(1S)$ and  $\Upsilon(2S)$ are observed. 
%The experimental challenge at the D0 experiment is the separation of the different Bottomium final states. 
Figure~\ref{fig:numbertwo}~(right) shows the measured differential cross section of the $\Upsilon(1S)$. Because of the high $\eta$ acceptance of the D0 muon system these measurement is done in three $y$ bins. No significant $y$ dependence is observed. %For the theoretical prediction a calculation using $k_T$ factorisation is used. 
It is planned to extend this analysis to a polarisation measurement using a data set with a luminosity of $1\,\mathrm{fb^{-1}}$. To extract the polarisation of all three resonances simultanously an unfolding method is necessary.  

\begin{figure}
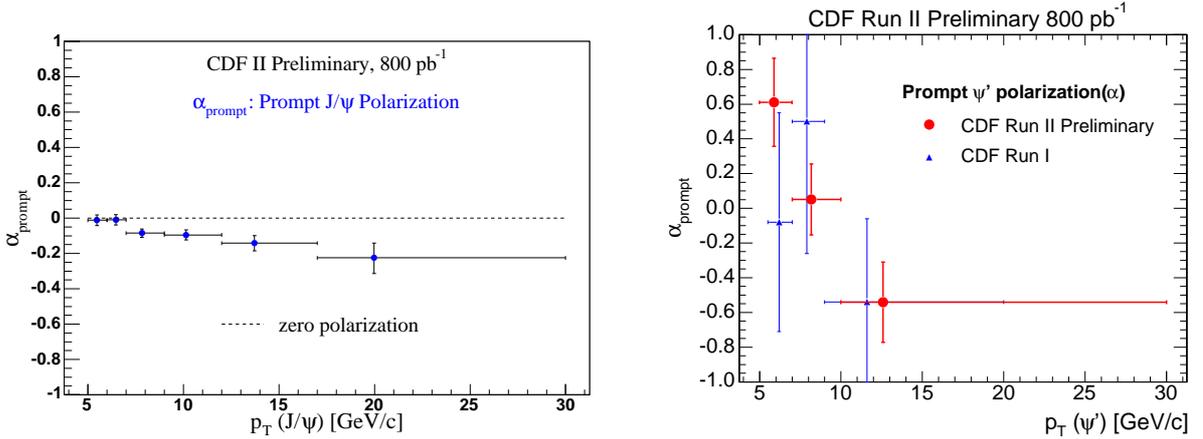

%\rule{5cm}{0.2mm}\hfill\rule{5cm}{0.2mm}
%\vskip 2.5cm
%\rule{5cm}{0.2mm}\hfill\rule{5cm}{0.2mm}
\epsfig{figure=polarization_alpha_prompt_jpsi_cmucmu_8_1-75.eps, width=0.54\textwidth}\epsfig{figure= psiprpol.eps, width=0.46\textwidth}
\caption{Polarisation ($\alpha$) of the vector meson as function of its $p_T$ for prompt $J/\Psi$ and $\Psi (2S)${ \label{fig:numberthree}}}
\end{figure}

The CDF collaboration measured  the polarisation for the two vector mesons $J/\Psi$ and $\Psi(2S)$~\cite{CDFJpsri1,CDFJpsri2} using an integrated luminosity of about $800\,\mathrm{pb^{-1}}$. Both modes were selected using the decay into two muons. The flight direction of the $\mu^+$ relative to the flight direction of the vector meson in the proton-proton rest frame of the $p\overline{p}$ measured by the polar angle $\theta^*$ which depends on the polarisation parameter $\alpha$ were $\alpha=1(-1)$ for transversal (longitudinal) polarisation. Figure~\ref{fig:numberthree} shows the results for prompt $J/\Psi$ (left) and the $\Psi(2S)$ (right) production. For $J/\Psi$ production the measured polarisation is longitudinal and significantly increases with the $p_T$. This agrees fully with the expectation of the new models. Also the  $\Psi(2S)$ polarisation shows a trend towards longitudinal polarisation with higher $p_T$.\\

Additionally the CDF collaboration measured the relative cross section of the $\chi_1$ and $\chi_2$ mesons~\cite{CDFJpsri3}. These mesons decay into $J/\Psi$ mesons via the radiation of a low energetic photon. The precission of the measurement of $\sigma(\chi_{c2})/\sigma(\chi_{c1}) = 0.70 \pm 0.04(\mathrm{stat.}) \pm 0.04(\mathrm{sys.}) \pm (0.06\,\mathrm{branching\,fractions})$ for the prompt production sets a new standard for this measurement. It excludes a naive estimation based on counting of the different spin orientations which leads to an expected cross-section ratio of 5:3.

\section{Summary} 
The study of heavy flavour production mechanism is an activ research topic at the Tevatron collider. Both experiments CDF and D0 have now collected a large sample of data with a luminosity of about $2\,\mathrm{fb^{-1}}$. With this huge data sample it will be possible to access the details of the production mechanism by the measurement of polarisation of quarkonium or the quark anti-quark correlation. This additional information is especially important because the theoretical knowledge of the cross section are dominated by scale uncertaincies. The new measurements provide an additional input information for the understanding of the production mechanism for open and hidden quark production in the transition region between pertubative and non-pertubative QCD.

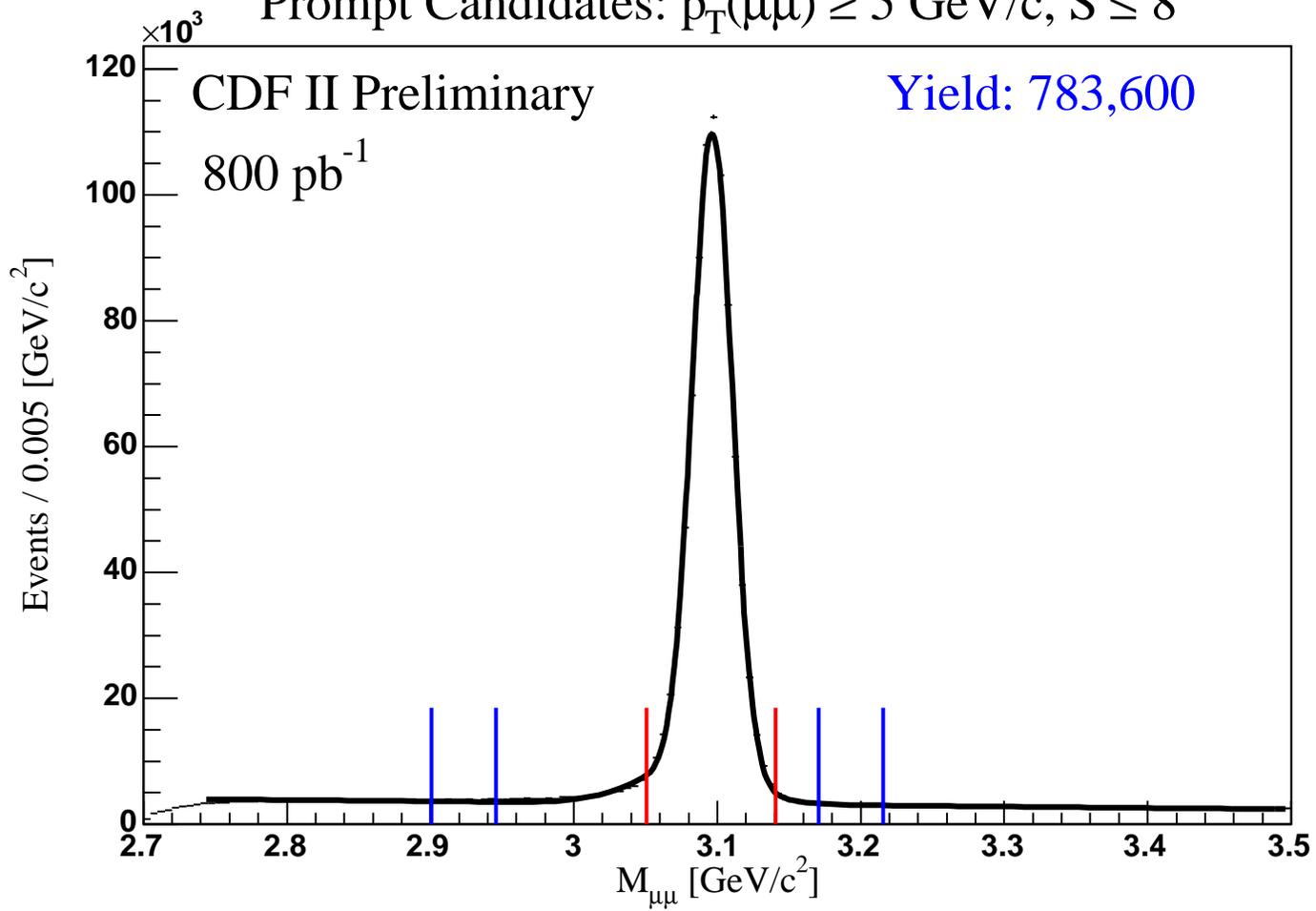
\end{document}